\documentclass[lettersize,journal]{IEEEtran}

\usepackage{times}  
\usepackage{helvet}  
\usepackage{courier}  
\usepackage[hyphens]{url}  
\usepackage{graphicx} 
\usepackage{cite} 
\usepackage{caption} 
\usepackage{algorithm}
\usepackage{algorithmic}

\usepackage{multirow}
\usepackage{booktabs}
\usepackage{amssymb}
\usepackage{amsmath}
\usepackage{subfigure}

\usepackage{subfigure}

\begin{document}





\title{Injection, Attack and Erasure: Revocable Backdoor Attacks via Machine Unlearning}

\author{\IEEEauthorblockN{Baogang Song,
Dongdong Zhao,
Jianwen Xiang, 
Qiben Xu and,
Zizhuo Yu
\thanks{This work was supported by the Hubei Province Major Science and Technology Innovation Program (2024BAA011). (Corresponding author: Dongdong Zhao, zdd@whut.edu.cn.)}%
\thanks{Baogang Song, Dongdong Zhao, Jianwen Xiang, Qiben Xu and, Zizhuo Yu are with the Engineering Research Center of Transportation Information and Safety (ERCTIS), MoE of China, School of Computer Science and Artificial Intelligence, Wuhan University of Technology, Wuhan, China}%
},



}

\markboth{Journal of \LaTeX\ Class Files,~Vol.~14, No.~8, August~2021}%
{Shell \MakeLowercase{\textit{et al.}}: A Sample Article Using IEEEtran.cls for IEEE Journals}


\maketitle

\begin{abstract}
Backdoor attacks pose a persistent security risk to deep neural networks (DNNs) due to their stealth and durability. While recent research has explored leveraging model unlearning mechanisms to enhance backdoor concealment, existing attack strategies still leave persistent traces that may be detected through static analysis. In this work, we introduce the first paradigm of revocable backdoor attacks, where the backdoor can be proactively and thoroughly removed after the attack objective is achieved. We formulate the trigger optimization in revocable backdoor attacks as a bilevel optimization problem: by simulating both backdoor injection and unlearning processes, the trigger generator is optimized to achieve a high attack success rate (ASR) while ensuring that the backdoor can be easily erased through unlearning. To mitigate the optimization conflict between injection and removal objectives, we employ a deterministic partition of poisoning and unlearning samples to reduce sampling-induced variance, and further apply the Projected Conflicting Gradient (PCGrad) technique to resolve the remaining gradient conflicts. Experiments on CIFAR-10 and ImageNet demonstrate that our method maintains ASR comparable to state-of-the-art backdoor attacks, while enabling effective removal of backdoor behavior after unlearning. This work opens a new direction for backdoor attack research and presents new challenges for the security of machine learning systems.
\end{abstract}

\section{Introduction}

Deep neural networks (DNNs) have achieved remarkable breakthroughs in recent years and are now widely used in critical domains such as image recognition and natural language processing. However, as DNN models become more prevalent, concerns about their security have also become increasingly prominent. Recent studies have demonstrated that DNN systems are susceptible to a variety of security threats in real-world scenarios, including adversarial examples \cite{ref1}, model stealing \cite{ref2}, and data poisoning. Among these, backdoor attacks \cite{ref3} have received considerable research attention due to their strong stealthiness and the significant harm they can cause.

In a typical backdoor attack, an adversary injects a small number of poisoned samples embedded with specific triggers into the training dataset during the model training process. Consequently, the resulting model performs normally on clean inputs but can be manipulated to output an attacker-chosen target label when a trigger is present in the input. Because triggers are usually designed to be highly covert and only activate the backdoor when present in the input, backdoor attacks are difficult to detect and defend against with conventional security methods. As researchers continue to develop more advanced backdoor detection and defense techniques, attackers are also exploring increasingly sophisticated and covert attack strategies, leading to an ongoing arms race in the field \cite{ref4}.

Recently, a novel attack paradigm has emerged. This paradigm \cite{ref5} exploits machine unlearning mechanisms to manipulate backdoor states. Originally, the unlearning mechanism was introduced to address privacy regulations (such as GDPR \cite{ref8}), with the goal of enabling models to completely eliminate the influence of specific data on their predictions. However, recent studies \cite{ref5,ref6,ref7} have revealed that attackers can cleverly abuse this mechanism: they can conceal backdoor effects and later reactivate backdoor behavior after model deployment by leveraging targeted unlearning operations.

\begin{figure}
    \centering
    \includegraphics[width=\columnwidth]{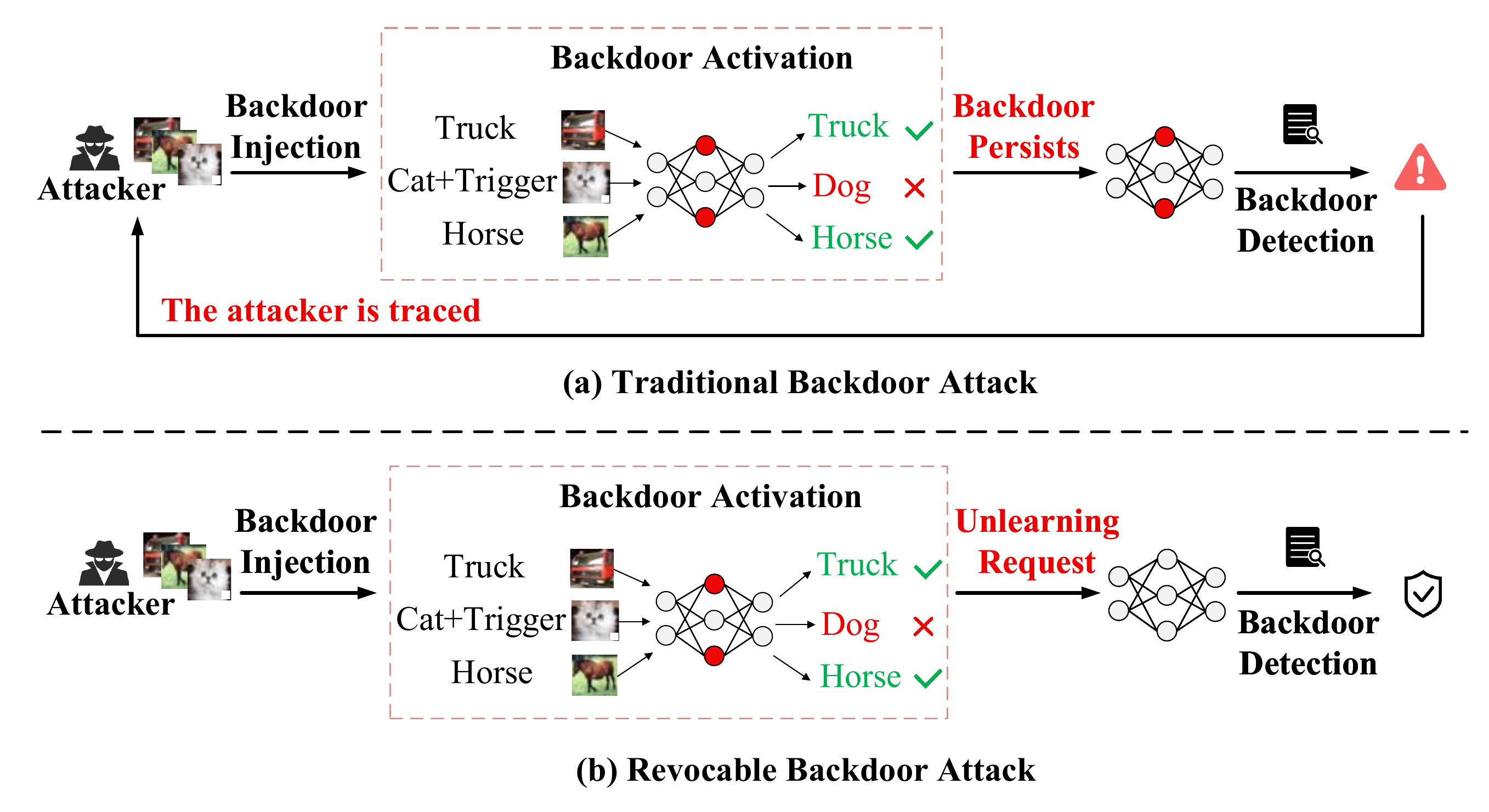}
    \caption{Research Motivation}
    \label{fig:motivation}
\end{figure}
Backdoor attacks based on unlearning mechanisms introduce the notion of delayed activation, where the model behaves normally at deployment and only activates the backdoor when triggered by a specific operation, thus greatly enhancing stealth. However, neither traditional backdoor attacks nor those utilizing unlearning mechanisms can eliminate the persistent risks that the long-term presence of backdoors poses for attackers. As illustrated in Figure~\ref{fig:motivation}, once a backdoor attack is carried out, the associated backdoor features remain embedded in the model’s parameter or activation space throughout the model’s lifecycle. Defenders can detect traces of backdoors at any stage of the model's lifecycle by using static analysis methods such as model pruning \cite{ref10}, trigger pattern inversion \cite{ref11}, and activation clustering \cite{ref9}. As a result, attackers can hardly avoid the risk of passive exposure. This situation raises an important but underexplored question: Can attackers actively revoke backdoors from a model after the attack is completed, thereby thoroughly erasing traces of the attack and evading long-term detection and tracking? Importantly, to achieve true stealthiness and practicality in real-world scenarios, such revocation should require unlearning only a minimal subset of samples, rather than resorting to large-scale data deletion, which could itself arouse suspicion during routine model management or compliance audits. If attackers can achieve this, it would significantly improve the stealthiness and risk avoidance of backdoor attacks, while also posing new challenges for current defense mechanisms:
\begin{enumerate}
    \item \textbf{Attribution difficulty}\enspace 
    Attackers could strategically remove critical samples after achieving their objectives or just before a model audit, causing the model to behave normally and fundamentally reducing the likelihood of attribution and accountability.
    \item \textbf{Static detection invalidation}\enspace 
    Most existing static backdoor detection methods \cite{ref9,ref10,ref11} assume that backdoor features persist in a model’s parameters or activation space. If backdoors can be actively revoked, these static detection tools become ineffective, forcing defenders to invest more resources in dynamic defenses such as continuous online monitoring and behavioral log auditing, to promptly detect and track backdoor attacks.
\end{enumerate}

To address this gap, we propose a novel paradigm of revocable backdoor attacks, which allows attackers to actively remove backdoors after achieving their goals, thereby thoroughly erasing attack traces and significantly enhancing stealth. Our approach adopts an alternating optimization framework: the trigger generator is trained to maximize attack success on a surrogate model while minimizing effectiveness after unlearning, ensuring the resulting triggers are both effective and easily revocable. To mitigate the optimization conflict between these objectives, we employ a deterministic sample partition during training and apply PCGrad to further reduce conflict and stabilize optimization. After training, the attacker uses the learned trigger generator to inject backdoors into the target model. Once the attack objective is reached, a forgetting request with clean labels is submitted, prompting unlearning to remove the backdoor. Experiments on CIFAR-10 and ImageNet show that our method maintains high primary task accuracy and attack success rates, while the backdoor can be significantly weakened or removed through unlearning, outperforming traditional backdoor attacks in stealth and revocability.

The main contributions are as follows:
\begin{itemize}
    \item We are the first to propose a revocable backdoor attack paradigm that leverages the model unlearning interface, enabling proactive backdoor removal and significantly enhancing attack stealth. This opens a new direction for backdoor attack research and highlights critical security challenges.
    \item We design an bilevel optimization-based trigger generator, jointly training surrogate and unlearning models to simulate poisoning and unlearning, and optimize trigger effectiveness and revocability.
    \item To address conflicts in trigger optimization, we introduce fixed poisoned/unlearning samples and employ the PCGrad technique, effectively stabilizing the optimization process.
    \item We evaluate the effectiveness of our method on datasets including CIFAR-10 and ImageNet. Results indicate that our method achieves attack success rates comparable to mainstream backdoor attacks, and that backdoor effects can be significantly weakened or removed after unlearning.
\end{itemize}
\section{Related Work}

\subsection{Backdoor Attacks}

Backdoor attacks have emerged as a significant threat to deep neural networks. Early works such as BadNets \cite{ref12} and Trojaning Attack \cite{ref14} demonstrated that models can be compromised by injecting poisoned samples containing simple triggers into the training set, achieving high attack success rates (ASR) on triggered inputs. Subsequent methods have improved the stealthiness, robustness, and transferability of triggers by adopting blended patterns \cite{ref13}, image warping \cite{ref16}, invisible triggers \cite{ref15}, or semantic triggers \cite{ref17}. More recent research \cite{ref5,ref6,ref7} has explored advanced backdoor strategies, such as delayed backdoors that exploit machine unlearning. However, existing approaches cannot proactively and thoroughly erase the backdoor; Attack traces often remain in the model parameters, which makes these traces susceptible to detection or tracing \cite{ref27, ref28}. Although there is a prior work \cite{ref29} that claims to achieve ``revocable'' backdoor attacks, its applicability is restricted to specific scenarios such as model trading, and it fundamentally differs from our proposed paradigm. In addition, this method requires modification of model parameters. Unlike our approach, which systematically enables proactive and generalizable backdoor removal through machine unlearning, the existing method does not address the broader and more practical challenge of actively erasing backdoors from the attacker's perspective.

\subsection{Machine Unlearning}

Machine unlearning was initially proposed to meet privacy and regulatory requirements such as GDPR, focusing on data removal and certified deletion in trained models. Mainstream unlearning methods include influence-based data removal \cite{ref30}, approximate retraining \cite{ref25, ref26}, and certified forgetting \cite{ref31}, which have shown effectiveness in erasing individual data points or entire subsets from a model's memory. Recently, the intersection between machine unlearning and security has drawn increasing attention. Some studies \cite{ref32} have explored leveraging unlearning mechanisms to mitigate backdoor attacks or defend against data poisoning. Nevertheless, these efforts primarily aim to enhance model robustness or privacy, rather than enabling attackers to actively control or erase backdoor traces. To the best of our knowledge, our work is the first to systematically exploit machine unlearning as a positive tool for active backdoor revocation from the attacker's perspective, achieving both attack effectiveness and removability.

\section{Revocable Backdoor Attack}
\begin{figure*}
    \centering
    \includegraphics[width=\textwidth]{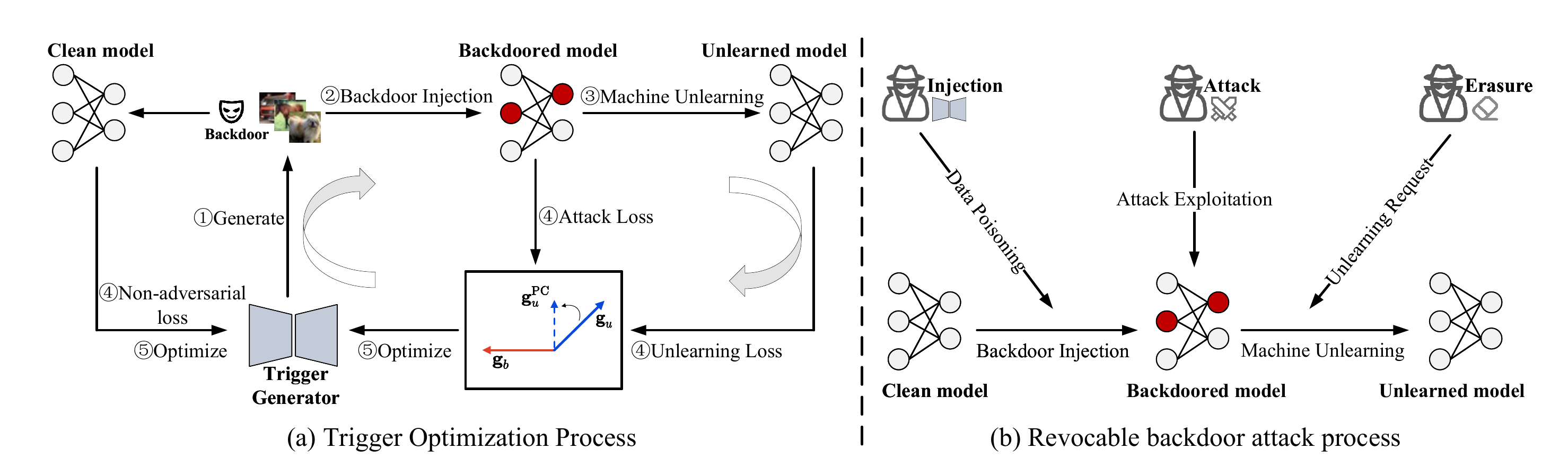}
    \caption{An illustration of the proposed framework}
    \label{fig:framework}
\end{figure*}

In this section, we first formally define the concept of a revocable backdoor attack and then elaborate on the trigger optimization process.
\subsection{Problem Definition}
As illustrated in Figure~\ref{fig:framework}, a revocable backdoor attack can be divided into three stages: trigger generation, backdoor injection, and backdoor revocation.
\begin{enumerate}
    \item \textbf{Trigger optimization:}\enspace 
    Unlike traditional backdoor attacks, our approach incorporates revocability as a core consideration during trigger optimization. Specifically, the trigger is designed not only to ensure a high ASR, but also to enable effective removal after model unlearning, thereby evading detection and long-term tracking.
    \item \textbf{Backdoor injection:}\enspace 
    Once the trigger is optimized, the attacker injects trigger samples into the training dataset. These samples are embedded in the victim model either through data upload or by submitting model training requests to the service provider.
    \item \textbf{Backdoor revocation:}\enspace 
    After achieving the attack goal, the attacker sends a forgetting request to the model provider, deletes some trigger-containing training data, and removes the backdoor via the legal model unlearning interface, thus hiding attack traces.
\end{enumerate}
We formally define the \textbf{Revocable Backdoor Attack} as the following optimization problem:
Given a classification model $f_\theta$ and its training dataset $\mathcal{D} = {(x_i, y_i)}$, where $\mathcal{X}$ and $\mathcal{Y}$ denote the input and label spaces, and $x_i \in \mathcal{X}$ and $y_i \in \mathcal{Y}$. The attacker's goal is to inject a backdoor into the model such that it can be reliably activated by a trigger, but subsequently revoked via an unlearning request. Specifically, the attacker first selects a target label $y_{\mathrm{target}} \in \mathcal{Y}$ and extracts a subset of samples with the target label, denoted as $\mathcal{P} \subset \mathcal{D}$, to serve as the poisoned set (with poisoning rate $\rho_p$). From this set, a further subset $\mathcal{U} \subset \mathcal{P}$ is sampled as the forgetting set (with forgetting rate $\rho_f$), which will be used for backdoor revocation.
Next, the attacker applies a trigger generator function $G$ to these samples, producing the poisoned set $\mathcal{P}^{\prime} = {(G(x), y) \mid (x, y) \in \mathcal{P}}$. The attacker then constructs a mixed training set $\mathcal{D}_b = (\mathcal{D} \setminus \mathcal{P}) \cup \mathcal{P}^{\prime}$ and submits it to the model owner for training. After training, the model parameters are updated to $\theta_b$, resulting in a backdoored model $f_{\theta_b}$, which behaves normally on clean inputs but predicts the target label $y_{\mathrm{target}}$ for inputs containing the trigger $G(x)$:
\begin{equation}
f_{\theta_b}(x_i)=y_i,\ f_{\theta_b}(G(x_i))=y_{target}\quad \forall x_i,y_i\in\mathcal{X},\mathcal{Y}
\end{equation}
After the attack objective is achieved, the attacker submits a forgetting request for the subset $\mathcal{U}$, prompting the model owner to invoke standard or approximate unlearning algorithms (such as fine-tuning, first-order or unrollsgd) to update the model parameters to $\theta_u$, thus obtaining the unlearned model $f_{\theta_u}$. If the trigger is no longer effective, the backdoor is successfully revoked, and traces of the attack are effectively erased:
\begin{equation}
f_{\theta_u}(x_i)=y_i,\ f_{\theta_u}(G(x_i))=y_i\quad \forall x_i,y_i\in\mathcal{X},\mathcal{Y}
\end{equation}

Therefore, our goal is to find a trigger generator $G$ such that, after training the model following the above workflow, the resulting models satisfy the following two properties: the trigger can reliably activate the backdoor before unlearning (Eq. 1), but loses its effect after unlearning (Eq. 2). The revocable backdoor attack is thus formulated as the problem of designing and optimizing $G$ to satisfy Eq. 1 and Eq. 2 simultaneously.

\subsection{Threat Model}
\begin{itemize}
    \item \textbf{Attacker's Capability}\enspace The attacker is able to inject or modify a small number of samples with backdoor triggers in the model’s training set but cannot interfere with the overall training process or the deployment of the final model. In addition, the attacker can utilize the data unlearning interface provided by the model service provider (e.g., for GDPR compliance) to request the model to forget a specified small subset of training samples. All poisoned samples must retain their true labels—that is, only clean-label backdoor attacks are considered. This clean-label setting is adopted for two main reasons: First, in practical scenarios, attackers typically lack the ability to alter data labels, making clean-label attacks more realistic. Second, it prevents exposure during the revocation phase, as requesting unlearning for mislabeled (dirty-label) samples would easily arouse suspicion from the service provider.
    \item \textbf{Attacker's Objective}\enspace The attacker aims to ensure that, when the trigger is present, the model predicts the attacker-specified target label with a high ASR. After the designated samples are forgotten, the model becomes fully insensitive to the trigger, which makes the attack difficult to detect or trace afterward. Unlike traditional backdoor attacks, which emphasize persistence, our work centers on \textbf{revocability}: the attacker seeks not only high ASR but also the ability to proactively and thoroughly erase all traces of the backdoor after achieving their goal, thereby significantly enhancing the stealth and safety of the attack.
\end{itemize}

\subsection{Trigger Generation and Optimization}

\subsubsection{Design of the Trigger Generator}
We define the trigger as an input perturbation function:
\begin{equation}
G(x)=x+\eta*g(x)
\end{equation}
where $g$ is a learnable generative network and $\eta$ represents the perturbation strength. To address high-frequency artifacts and spatial discontinuities \cite{ref18}, we incorporate design principles inspired by the COMBAT method \cite{ref19}. Specifically, we first apply the discrete cosine transform (DCT) \cite{ref20} to $g(x)$ to constrain the trigger's frequency domain distribution. A mask $m$ is then used to control the retained frequency components. Subsequently, we reconstruct the trigger noise in the spatial domain using the inverse DCT (IDCT), resulting in a trigger that is dominated by low-frequency components. To further enhance spatial continuity and stealth, we apply a Gaussian blur filter $k$ to the generated trigger. By integrating these steps, we obtain a highly covert trigger function for backdoor attacks. The final trigger function is defined as:
\begin{equation}
    \mathcal{G}(g(x))=\mathrm{IDCT}(m\odot\mathrm{DCT}(g(x)))
\end{equation}
\begin{equation}
    G(x)=(x+\eta*\mathcal{G}(g(x)))*k
\end{equation}
where $\odot$ denotes Hadamard product.

\subsubsection{Bilevel Optimization}
In revocable backdoor attacks, the core challenge lies in the fact that the effectiveness of the trigger depends not only on the model's behavior after backdoor injection, but also on its behavior after subsequent unlearning operations. Unlike conventional backdoor attacks, which focus solely on attack effectiveness during the backdoor injection phase, revocable backdoor optimization must fully account for the model's state after unlearning. Therefore, when optimizing the trigger, we explicitly simulate both the backdoor injection and unlearning processes to ensure that the trigger achieves its intended objectives at each stage. This necessity motivates us to formulate the training of the trigger generator as a bilevel optimization problem. Specifically, in the bilevel optimization framework, the \textbf{outer level} optimizes the parameters of the trigger generator, while the \textbf{inner level}, given the current trigger generator, sequentially trains the backdoored model and performs the unlearning operation, thereby providing optimization signals from both the backdoor injection and unlearning phases. It is important to note that in real-world attack settings, the attacker is typically unable to access the internals of the victim’s deployed model or intervene in its training process. To address this limitation, we adopt a surrogate training approach, in which a locally constructed shadow model is built to match the architecture of the victim model as closely as possible. We then utilize auxiliary data accessible to the attacker to simulate the processes of backdoor injection and unlearning on this shadow model.

In the outer-level trigger optimization, we focus on two core loss terms:
\begin{itemize}
    \item \textbf{Attack loss}: In the outer-level trigger optimization, the attack loss is computed by applying the trigger to all samples in the dataset, assigning them the target label $y_{target}$, and feeding the triggered samples into the backdoored model for evaluation:
     \begin{equation}
        \mathcal{L}_{attack} = \mathbb{E}_{(x, y) \in \mathcal{D}}\, \mathcal{L}\left(f_{\theta_b}(G(x)),\, y_{target}\right)
    \end{equation}
    This design aims to encourage the learned trigger to be generally applicable, rather than limited to a specific subset of samples.
     \item \textbf{Unlearning loss}: Similarly, the unlearning loss is computed by applying the trigger to all samples and evaluating the triggered samples with the unlearned model, using the ground-truth labels $y$ for comparison:
    \begin{equation}
        \mathcal{L}_{unlearn} = \mathbb{E}_{(x, y) \in \mathcal{D}}\, \mathcal{L}\left(f_{\theta_u}(G(x)),\, y\right)
    \end{equation}
    This loss encourages the complete revocation of the backdoor effect, i.e., the unlearned model should no longer respond to the trigger.
\end{itemize}
These two losses jointly define the main objectives of trigger optimization, i.e., effectiveness and revocability. Furthermore, we also employ the following two losses:
\begin{itemize}
    \item \textbf{Visibility loss.} To ensure the trigger remains visually inconspicuous, we penalize visually salient perturbations using the following regularization:
\begin{equation}
    \mathcal{L}_{vis} = \mathbb{E}_{x\in \mathcal{X}}\left[|\eta*\mathcal{G}(g(x))|_2^2\right]
\end{equation}
    \item \textbf{Non-adversarial loss.} To prevent the trigger from acting as a standard adversarial perturbation and to maintain the specificity of the backdoor, we require that the clean model makes consistent predictions on triggered and clean samples:
\begin{equation}
    \mathcal{L}_{non\text{-}adv} = \mathbb{E}_{(x, y) \in \mathcal{D}} \ \mathcal{L}\left(f_{\theta_{clean}}(G(x)),\, y\right)
\end{equation}
\end{itemize}
Combining the above objectives, the trigger generator is trained to solve the following bilevel optimization problem:
\begin{align}
G^* =\ & \arg\min_G\, \mathbb{E}_{(x, y) \in \mathcal{D}} \left\{
\begin{array}{l}
     \mathcal{L}_{attack}  +\, \lambda_{unlearn}\cdot\mathcal{L}_{unlearn} \\[2pt]
    +\, \lambda_{vis}\cdot\mathcal{L}_{vis} \\[2pt]
    +\, \lambda_{non\text{-}adv}\cdot\mathcal{L}_{\mathrm{non\text{-}adv}}
\end{array}
\right. \\
\text{s.t.}\quad
& \theta_b = \arg\min_{\theta}\,  \mathbb{E}_{(x, y) \in \mathcal{D}_b} \mathcal{L}\left(f_{\theta}(x),\, y\right), \notag \\
& \theta_u = \mathcal{F}_{unlearn}(\theta_b,\, \mathcal{U}) \notag
\end{align}
where $\lambda_{unlearn}$, $\lambda_{vis}$, and $\lambda_{non\text{-}adv}$ are the weights for the unlearning, visibility, and non-adversarial losses, respectively. 
$\mathcal{F}_{unlearn}(\theta_b,\, \mathcal{U})$ denotes the model unlearning algorithm that updates the backdoored model parameters $\theta_b$ by removing the influence of the specified forgetting set $\mathcal{U}$.

Due to the strong coupling and mutual dependencies among parameters in the above optimization problem, direct end-to-end optimization of the overall objective is difficult to converge. Specifically, the training of the backdoored and unlearned models depends on the poisoned data generated by the current trigger generator parameters, while the optimization of the trigger generator in turn relies on feedback from both models. This circular dependency results in a highly complex parameter space and interdependent objectives. Therefore, we adopt an alternating optimization strategy: in each training round, we first fix the trigger generator parameters to train the backdoored and unlearned models, and then fix the model parameters to optimize the generator.

\subsubsection{Gradient Conflict Mitigation}
In the bilevel optimization process, maximizing the ASR of the backdoored model and minimizing the ASR of the unlearned model are inherently conflicting objectives: increasing the backdoored model's ASR with respect to the trigger often makes it more difficult for the unlearned model to revoke the backdoor, and vice versa. This conflict manifests as inconsistency in the directions of the gradients of the loss functions during optimization; in severe cases, the gradients may even cancel each other out, undermining the stability and final performance of trigger generator training. Empirically, we observe that the gradient directions of these two losses often exhibit significant negative correlation during training. To effectively alleviate such gradient conflicts between optimization objectives, we adopt the following two mechanisms:
\begin{enumerate}
    \item \textbf{Deterministic partition:}\enspace 
    In practical backdoor attacks, adversaries typically select a specific set of samples for poisoning, which aligns with our use of a deterministic partition. While many existing methods simulate backdoor injection by randomly sampling data during optimization in order to generate triggers that generalize across the entire dataset, we instead fix the sample partition throughout the optimization process. This approach more faithfully reflects realistic attack settings, reduces sampling-induced variance, and allows for more stable optimization of the trigger generator.
    \item \textbf{PCGrad gradient projection:}\enspace 
    When the gradient directions of the two loss functions conflict (i.e., their inner product is negative), we employ the PCGrad (Projected Conflicting Gradient) algorithm \cite{ref21} to orthogonally project the gradients and remove conflicting components. Specifically, let the gradients of the attack loss and unlearning loss with respect to the generator $G$ be denoted as $\mathbf{g}_b$ and $\mathbf{g}_u$, respectively:
\begin{equation}
\mathbf{g}_b = \nabla_G \mathcal{L}_{attack}
\end{equation}
\begin{equation}
\mathbf{g}_u = \nabla_G \mathcal{L}_{unlearn}
\end{equation}
When $\langle \mathbf{g}_b, \mathbf{g}_u \rangle < 0$, we project $\mathbf{g}_u$ onto the orthogonal direction of $\mathbf{g}_b$ as follows:
\begin{equation}
\begin{aligned}
\mathbf{g}_u^\mathrm{PC} & =\mathbf{g}_u-\alpha\cdot\frac{\langle\mathbf{g}_b,\mathbf{g}_u\rangle}{\|\mathbf{g}_b\|^2}\mathbf{g}_b
\end{aligned}
\end{equation}
where $\alpha$ is a hyperparameter controlling the projection magnitude and $\langle \cdot, \cdot \rangle$ denotes the standard inner product between two vectors.

\end{enumerate}

\begin{table*}[h]
    \centering
   
    \scalebox{1}{
    \begin{tabular}{c c c c c c | c c c c}
    \toprule
    \multirow{2}{*}{\textbf{Dataset}} & \multirow{2}{*}{\textbf{Unlearn Method}} & \multicolumn{4}{c|}{\textbf{First-Order}} & \multicolumn{4}{c}{\textbf{UnrollSGD}} \\
    \cline{3-10}
    \addlinespace
     &  & \textbf{ASR} & \textbf{ASR-U}($\Delta$) & \textbf{BA} & \textbf{BA-U} & \textbf{ASR} & \textbf{ASR-U}($\Delta$) & \textbf{BA} & \textbf{BA-U}\\
    \midrule
       \multirow{2}{*}{\textbf{CIFAR-10}}  & First-Order & 98.36 & 25.12(-73.24)  & 94.34 & 90.05 & 98.36 & 4.42(-93.94)  & 94.34 & 82.93 \\
         & UnrollSGD   & 97.33 & 12.93(-84.40)  & 94.21 & 91.03 & 97.33 & 0.11(-97.22)  & 94.21 & 83.29 \\
      \midrule
       \multirow{2}{*}{\textbf{IMAGENET-10}} & First-Order & 79.33    & 13.70(-65.63)     & 85.92    & 78.25    & 79.33    & 0.00(-79.33)     & 85.92    & 71.67    \\  
        & UnrollSGD   & 85.19    & 16.85(-68.34)     & 85.08    & 78.92    & 85.19    & 7.04(-78.15)     & 85.08    & 72.42    \\  
    \bottomrule
    \end{tabular}
    }
     \caption{Performance of Our Method under Different Unlearning Strategies on CIFAR-10 and ImageNet-10
 ($\%$)}
     \label{table:table1}

\end{table*}

\begin{table*}[h]
    \centering
    \scalebox{1}{
    \begin{tabular}{c c c c c | c c c c}
    \toprule
    \multirow{2}{*}{\textbf{Method}} & \multicolumn{4}{c}{\textbf{First-Order}} & \multicolumn{4}{c}{\textbf{UnrollSGD}} \\
    
    \cline{2-9}
    \addlinespace

     & \multirow{1}{*}{\textbf{ASR}} & \multirow{1}{*}{\textbf{ASR-U}}($\Delta$) & \multirow{1}{*}{\textbf{BA}} & \multirow{1}{*}{\textbf{BA-U}} & \multirow{1}{*}{\textbf{ASR}} & \multirow{1}{*}{\textbf{ASR-U}}($\Delta$) & \multirow{1}{*}{\textbf{BA}} & \multirow{1}{*}{\textbf{BA-U}}\\

    \midrule
     Badnets & 29.8 & 6.67(-23.13) & 94.47 & 92.19 & 29.8 & 1.12(-28.68)  & 94.47 & 89.33 \\
       Wanet & 65.28 & 29.51(-35.77)  & 90.45 & 89.54 & 65.28 & 6.41(-58.87)  & 90.45 & 89.81 \\
       Sleeper Agent & 90.26 & 57.47(-32.79)  & 93.96 & 91.89 & 90.26 & 18.57(-71.73)  & 93.96 & 83.56 \\
       COMBAT & \textbf{98.53} & 45.97(-52.56)  & 94.38 & 90.38 & \textbf{98.53} & 17.56(-80.97)  & 94.38 & 82.28 \\
       Our Method (First-Order) & \underline{98.36} & \underline{25.12(-73.24)}  & 94.34 & 90.05 & \underline{98.36} & \underline{4.42(-93.94)}  & 94.34 & 82.93 \\
       Our Method (UnrollSGD) & 97.33 & \textbf{12.93(-84.40)}  & 94.21 & 91.03 & 97.33 & \textbf{0.11(-97.22)}  & 94.21 & 83.29 \\
    \bottomrule
    \end{tabular}
    }
    \caption{
Comparison with state-of-the-art methods on CIFAR-10 (\%). 
\textbf{Bold} indicates the best result, and \underline{underline} indicates the second-best result in each column.
}
    \label{table:table2}

\end{table*}

\section{Experiment}

\subsection{Experimental Setup}

\subsubsection{Datasets and Models}
We evaluate the proposed method on two benchmark image classification datasets: \textbf{CIFAR-10} \cite{ref22} and \textbf{IMAGENET-10}. IMAGENET-10 is constructed by randomly selecting 10 classes from the standard ImageNet \cite{ref23} dataset. For model architectures, we use the Pre-activation ResNet-18 \cite{ref33} for CIFAR-10 and the standard ResNet-18 \cite{ref34} for IMAGENET-10. The trigger generator is implemented with a U-NET \cite{ref35} backbone in all experiments.

\subsubsection{Parameter and Attack Settings}
All classifiers are trained for 200 epochs with SGD. The batch size is 128 for CIFAR-10 and 32 for IMAGENET-10. Initial learning rates are 0.01 for CIFAR-10, 0.001 for IMAGENET-10, and 0.01 for the trigger generator, with decay by a factor of 10 at the 100th and 150th epochs. Class 0 is used as the target label in all attack experiments. The poisoning rate is fixed at 5\% in all experiments. For high-frequency removal, following COMBAT, we set the frequency mask ratio $r = 0.65$ and apply a Gaussian blur filter (kernel size 3, $\sigma$ uniformly sampled from $[0.1, 1]$). The hyperparameters $\alpha$, $\lambda_{\mathrm{non\text{-}adv}}$, $\lambda_{\mathrm{unlearn}}$, and $\lambda_{\mathrm{vis}}$ are set to 0.6, 0.8, 1.0, and 0.02, respectively. The PCGrad projection coefficient is fixed at 0.6. Unlearning is simulated using both the First-Order \cite{ref25} and UnrollSGD \cite{ref26} approximation methods. For each unlearning operation, 250 samples are removed from CIFAR-10 and 100 samples from IMAGENET-10. The forgetting rate for the First-Order setting is 0.01 (CIFAR-10) and 0.001 (IMAGENET-10); for UnrollSGD, CIFAR-10 is fine-tuned for 3 epochs and IMAGENET-10 for 2 epochs. Unless otherwise specified, all experiments are conducted on CIFAR-10, using the first-order unlearning strategy for both simulation and evaluation.

\subsubsection{Baselines and Evaluation Metrics}
To the best of our knowledge, there is currently no public research specifically targeting revocable backdoor attacks. For comprehensive comparison, we introduce several mainstream backdoor attack baselines (including BadNets \cite{ref12}, WaNet \cite{ref16}, Sleeper Agent \cite{ref24}, and COMBAT \cite{ref19}), and evaluate their performance changes after unlearning under the same settings. The evaluation metrics used are as follows: \textbf{Attack Success Rate (ASR)}, which measures the proportion of trigger samples classified as the target label; \textbf{ASR after Unlearning (ASR-U)}; \textbf{Benign Accuracy (BA)}, the classification accuracy on clean test samples; and \textbf{Benign Accuracy after Unlearning (BA-U)}.

\subsection{Attack Experiments}
We evaluate our method under First-Order and UnrollSGD unlearning on CIFAR-10 and ImageNet-10. As shown in Table~\ref{table:table1}, our approach consistently achieves high ASR and BA, with ASR-U dropping sharply after unlearning—especially under UnrollSGD, where ASR-U falls to 4.42\% on CIFAR-10 and 0.00\% on ImageNet-10, while BA-U remains competitive. Furthermore, as shown in Table~\ref{table:table2}, our comparison with Badnets, Wanet, Sleeper Agent, and COMBAT on CIFAR-10 demonstrates that our method not only achieves comparable or higher ASR but also yields substantially lower ASR-U after unlearning, particularly under UnrollSGD (0.11\%), highlighting its superior revocability. The detailed experimental settings for all compared methods are provided in the Appendix. These results confirm that our approach effectively balances attack effectiveness and removability, and outperforms existing methods in terms of revocable backdoor attacks.

\subsection{Defense Experiments}
\begin{figure}[h]
    \centering
    \subfigure{
        \includegraphics[width=0.45\columnwidth]{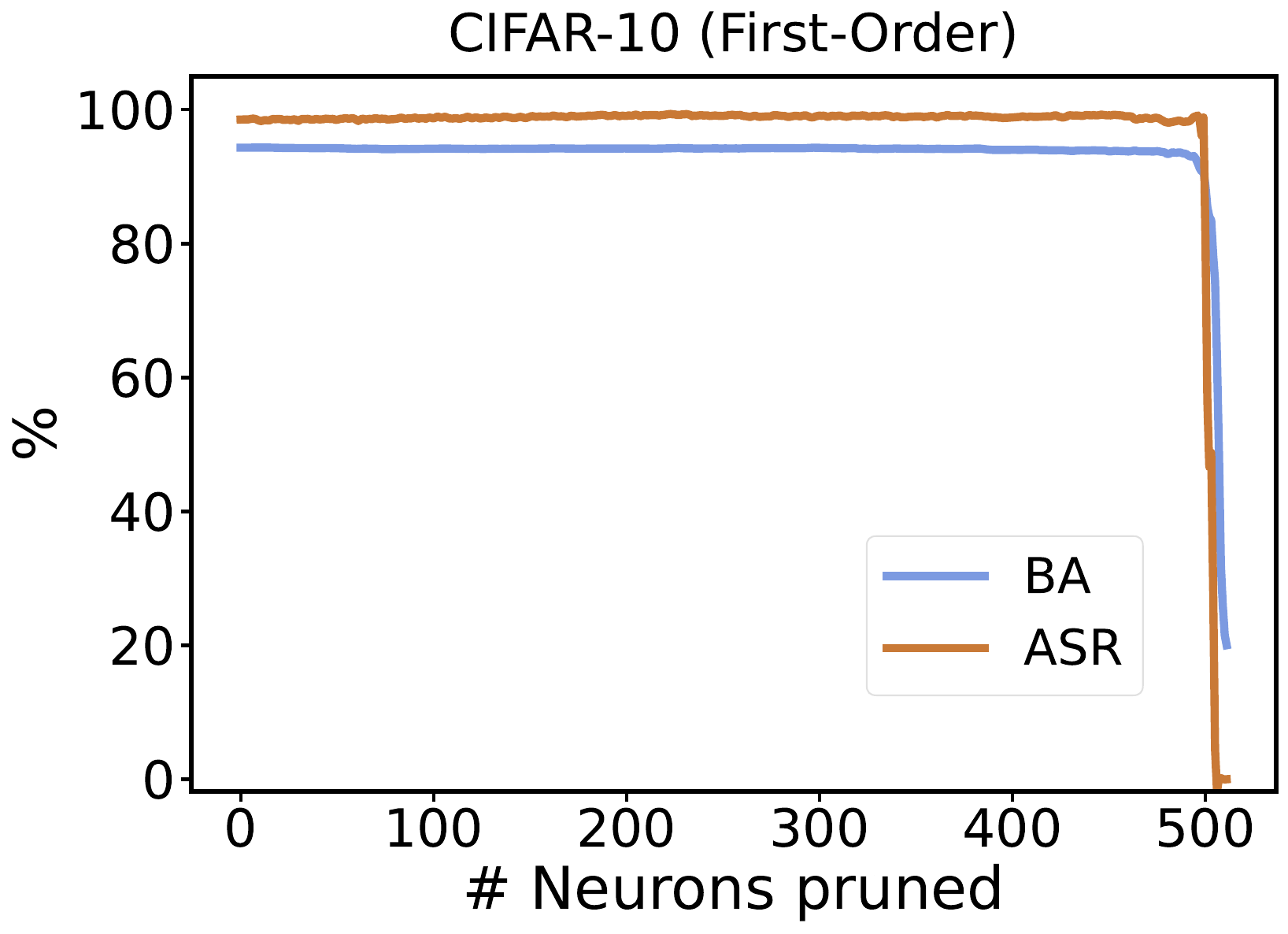}
    }
    \hfill
    \subfigure{
        \includegraphics[width=0.45\columnwidth]{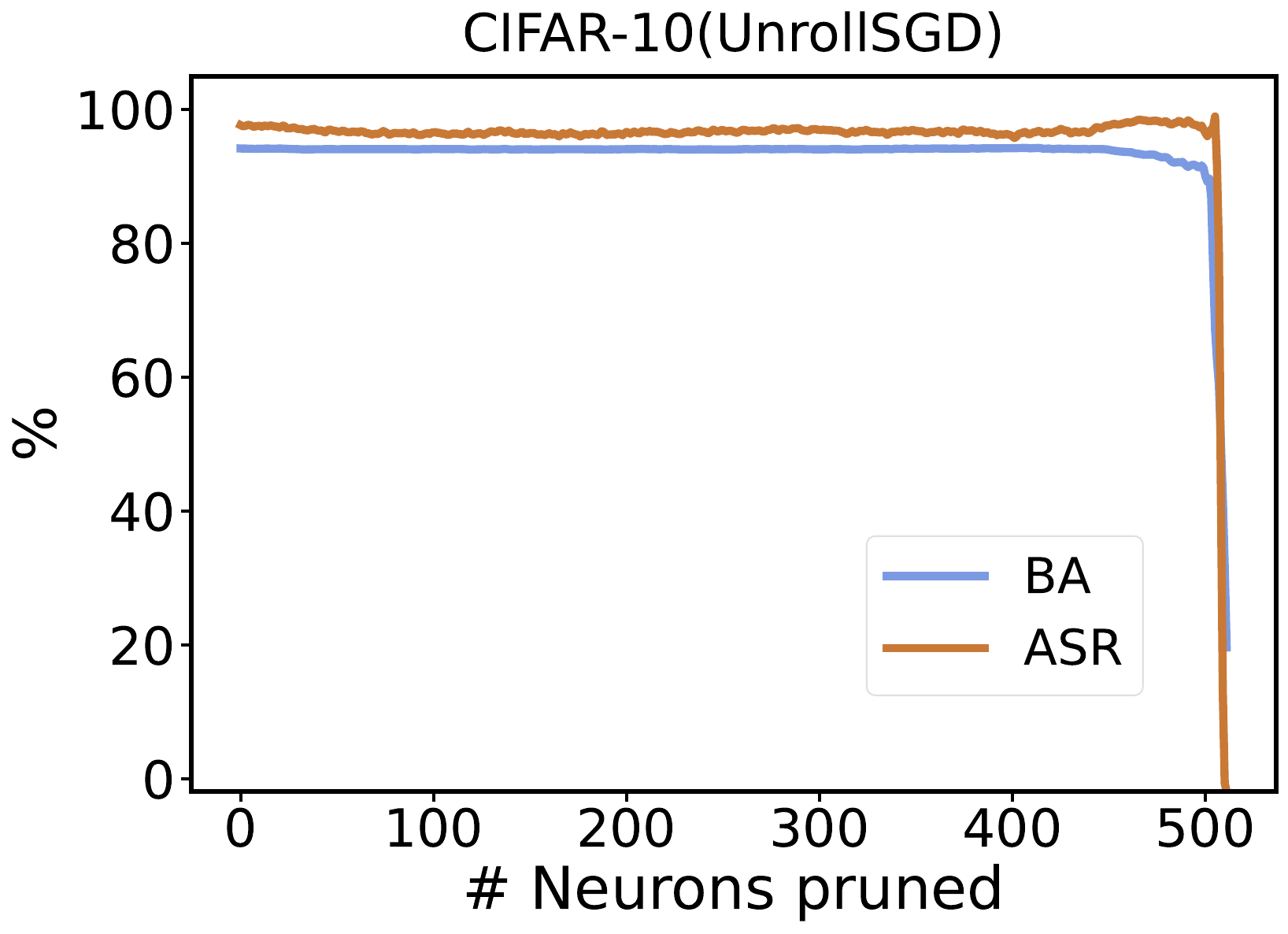}
    }
    \caption{Experiment results of evaluating Our Method against  Fine-pruning}
    \label{fig:defense}
\end{figure}

We evaluate our revocable backdoor using the fine-pruning defense, which prunes neurons inactive on clean samples. As shown in Fig.~\ref{fig:defense}, both BA and ASR remain high across a wide range of pruning. Only when a large number of neurons are pruned do both BA and ASR drop sharply, indicating that fine-pruning cannot effectively remove our backdoor without significantly damaging normal model performance. We further evaluate additional mainstream defenses on both CIFAR-10 and ImageNet-10, with detailed results provided in the Appendix.

\subsection{Parameter Sensitivity Analysis}
We conduct parameter sensitivity analysis on two hyperparameters: poisoning rate $\rho_p$ and trigger strength $\eta$. As shown in Table~\ref{tab:sensitivity_poison}, increasing $\rho_p$ leads to higher ASR and ASR-U, while BA and BA-U remain stable. This indicates that a higher poisoning rate strengthens the attack but also makes the backdoor more difficult to remove via unlearning. Similarly, Table~\ref{tab:sensitivity_etar} shows that increasing $\eta$ rapidly saturates ASR and substantially increases ASR-U, reducing revocability. When $\eta \geq 0.12$, further increases have little effect, indicating performance saturation. Across all settings, BA and BA-U remain largely unaffected. Overall, these results demonstrate a trade-off: while stronger triggers and higher poisoning rates improve attack effectiveness, excessive values do not yield further benefits and hinder unlearning-based backdoor revocation.

\begin{table}[h]
\centering
\begin{tabular}{ccccc}
\toprule
$\rho_p$($\%$)  & \textbf{ASR} & \textbf{ASR-U}($\Delta$) & \textbf{BA} & \textbf{BA-U} \\
\midrule
0.5   & 69.73 & 5.34(-64.39) & 94.76 & 90.81 \\
1   & 71.98 & 7.62(-64.36) & 94.38 & 90.74\\
2   & 84.38 & 12.19(-72.19) & 94.76 & 91.09\\
3   & 89.93 & 13.00(-76.93) & 94.50 & 90.75\\
5   & 98.36 & 25.12(-73.24) & 94.34 & 90.05\\
\bottomrule
\end{tabular}
\caption{Performance under different $\rho_p$ on CIFAR-10}
\label{tab:sensitivity_poison}
\end{table}

\begin{table}[h]
\centering
\begin{tabular}{ccccc}
\toprule
$\eta$  & \textbf{ASR} & \textbf{ASR-U}($\Delta$) & \textbf{BA} & \textbf{BA-U} \\
\midrule
0.04   & 71.82 & 9.71(-62.11) & 94.52 & 90.03 \\
0.08   & 98.36 & 25.12(-73.24) & 94.34 & 90.05\\
0.12   & 99.91 & 44.40(-55.51) & 94.41 & 90.82\\
0.16   & 99.81 & 41.24(-58.57) & 94.40 & 90.85\\
0.2  & 99.82 & 40.08(-59.74) & 94.33 & 90.61\\
\bottomrule
\end{tabular}
\caption{Performance under different $\eta$ on CIFAR-10}
\label{tab:sensitivity_etar}
\end{table}


\subsection{Gradient Conflict Analysis}
\begin{figure}[h]
    \centering
    \includegraphics[width=0.9\columnwidth]{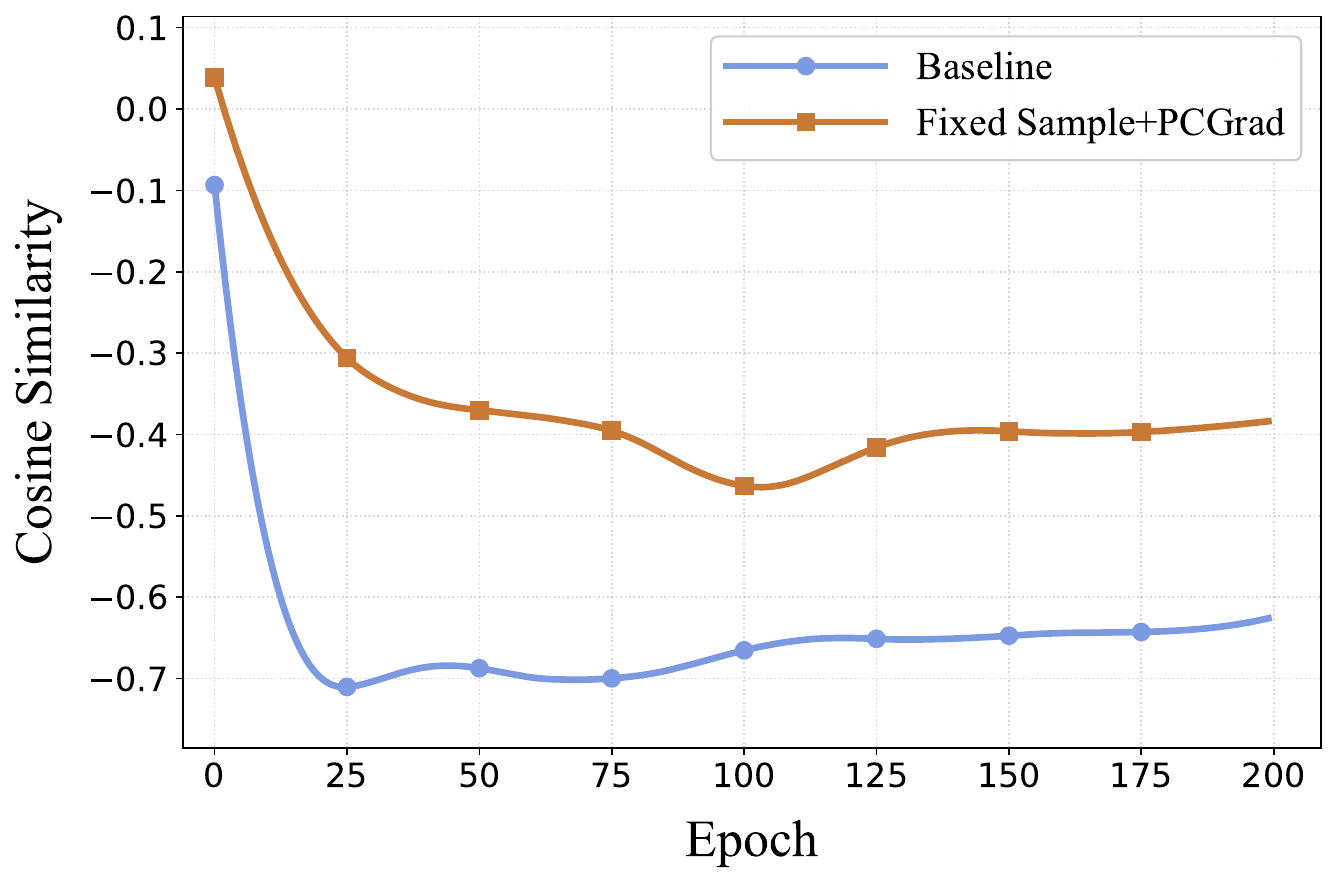}
    \caption{Cosine similarity between attack and unlearning gradients over training epochs}
    \label{fig:grad}
\end{figure}

To quantitatively investigate the optimization conflict between the backdoor injection and unlearning objectives, we measure the cosine similarity between the gradients of the attack loss and unlearning loss with respect to the trigger generator parameters during training. Here, cosine similarity is adopted as a quantitative indicator of gradient conflict, with lower values reflecting stronger opposition between objectives. We compare two settings: (1) \textbf{Baseline}, which does not employ any conflict mitigation strategy, and (2) \textbf{Fixed Sample + PCGrad}, which adopts both fixed sample selection and the PCGrad projection method to alleviate gradient conflict. As shown in Fig.~\ref{fig:grad}, the Baseline setting maintains a consistently strong negative cosine similarity ($-0.65$), indicating persistent and severe optimization conflict throughout training. In contrast, the Fixed Sample + PCGrad strategy achieves a significantly higher and more stable cosine similarity ($-0.35$), reflecting a substantial reduction in gradient conflict. These results demonstrate that our conflict mitigation approach effectively alleviates the opposition between the two objectives, resulting in a more stable and efficient optimization process for the training of revocable backdoor attack. The impact of gradient conflict mitigation on attack effectiveness and revocability will be further evaluated in the subsequent ablation study.

\subsection{Ablation Study}

To assess the contribution of each core module in our framework, we conduct ablation experiments by selectively removing the unlearning loss and the conflict mitigation strategy (Fixed Sample + PCGrad). The results on CIFAR-10 are summarized in Table~\ref{tab:ablation}. As shown, omitting the unlearning loss (w/o Unlearn) leads to a clear increase in ASR-U, indicating that the backdoor cannot be effectively revoked. Disabling the conflict mitigation strategy (w/o Mitigation) further exacerbates this issue, resulting in higher ASR-U. In contrast, our complete method (Ours) achieves the best balance, with high ASR, the lowest ASR-U, and BA comparable to other settings. These results confirm that both the unlearning loss and the conflict mitigation module are essential for achieving high attack effectiveness and strong revocability in revocable backdoor attacks.

\begin{table}[h]
\centering
\begin{tabular}{ccccc}
\toprule
\textbf{Setting}              & \textbf{ASR} & \textbf{ASR-U}($\Delta$) & \textbf{BA} & \textbf{BA-U} \\
\midrule
w/o Unlearn           &   97.50            &   32.52(-64.98)              &  94.58    & 90.59       \\
w/o Mitigation     &   98.57            &   68.60(-29.97)              &  94.51     & 91.34     \\
Ours                          &   98.36            &    25.12(-73.24)              &  94.34    & 90.05      \\
\bottomrule
\end{tabular}
\caption{Ablation study of key modules on CIFAR-10}
\label{tab:ablation}

\end{table}

\section{Conclusion}
In this paper, we propose a revocable backdoor attack paradigm that leverages machine unlearning to proactively and thoroughly erase backdoor traces. Our bilevel optimization-based trigger generator balances attack effectiveness and revocability, while practical techniques such as clean-label poisoning, fixed sample selection, and gradient conflict mitigation promote stability and stealth. Experiments on CIFAR-10 and ImageNet show that our method achieves attack success rates on par with state-of-the-art attacks, while enabling efficient backdoor removal via unlearning and surpassing baselines in both stealth and revocability. This work highlights new challenges for backdoor defense. In future work, we will explore more general attack and defense scenarios, including adaptive unlearning strategies, adversarial model auditing, and new defenses against revocable backdoor attacks.

 

\bibliographystyle{IEEEtran}
\bibliography{aaai2026}

\clearpage
\appendices

\section{Additional Implementation Details}

This appendix provides further technical details and supplementary analyses to facilitate reproducibility and a deeper understanding of our experimental methodology.


\subsection{Datasets and Preprocessing}

In all our comparative experiments, we evaluate every method on the \textbf{CIFAR-10} and \textbf{IMAGENET-10} datasets. For fairness, all methods use \emph{identical data splits and preprocessing pipelines}. CIFAR-10 consists of 50,000 training images and 10,000 test images, covering 10 object categories. IMAGENET-10 is constructed by randomly selecting 10 mutually exclusive classes from the standard ImageNet dataset, following the protocol described in the main text. All images are resized to $32\times32$ for CIFAR-10 and $224\times224$ for IMAGENET-10. For both datasets, we apply standard normalization using the per-channel mean and standard deviation. Data augmentation includes random horizontal flipping, random rotation and random cropping for CIFAR-10, and random resized cropping for IMAGENET-10. 

\subsection{Poisoned Sample and Unlearning Sample Selection}

For all compared methods, we adopt the \emph{clean-label backdoor attack} setting, in which only clean (i.e., correctly labeled) samples are selected for poisoning. Specifically, both poisoned samples and unlearning samples are consistently chosen from images belonging to class label $0$ across all evaluated approaches.

To ensure fair comparison, the exact same set of images from label $0$ is used as poisoned samples and as unlearning samples for all methods, unless otherwise stated. This fixed sample selection eliminates randomness and facilitates direct comparison of unlearning effectiveness.

\textbf{Exception:} The \emph{Sleeper Agent} method~\cite{ref24} employs a dynamic sampling strategy, where poisoned and unlearning samples are selected adaptively during training. As a result, for this method we do not fix the poisoned or unlearning sample set, in line with its original protocol.

\subsection{Unlearning Methods Adopted}

In our experiments, we utilize two unlearning strategies: \textbf{first-order unlearning} and \textbf{UnrollSGD unlearning}. Both are designed to effectively remove the backdoor association from the trained model, while minimizing any negative impact on its clean accuracy.

\subsubsection{First-order unlearning} is a straightforward method that applies a small number of gradient descent steps on the set of poisoned or unlearning samples. This approach is efficient and easy to implement, relying only on standard first-order optimization.

\subsubsection{UnrollSGD unlearning}, on the other hand, simulates a more advanced unlearning process by unrolling several steps of stochastic gradient descent on the unlearning samples. This allows the model to more thoroughly erase traces of the backdoor, at the cost of higher computational complexity.

To ensure model stability after unlearning, we only update the parameters of the last 15 layers for both the ResNet-18 and pre-activation ResNet-18 architectures during the unlearning process, keeping all earlier layers fixed.

\subsection{Defense Experiments}

In this section, we systematically evaluate our proposed backdoor attack against several widely used defenses: Fine-Pruning, STRIP, and GradCAM-based inspection. All evaluations are conducted on both CIFAR-10 and IMAGENET-10 datasets, considering both first-order and UnrollSGD unlearning strategies. Each defense is applied following the original protocol.
\subsubsection{Fine-Pruning}
\begin{figure*}[t]
  \centering
  \subfigure{
    \includegraphics[width=0.32\linewidth]{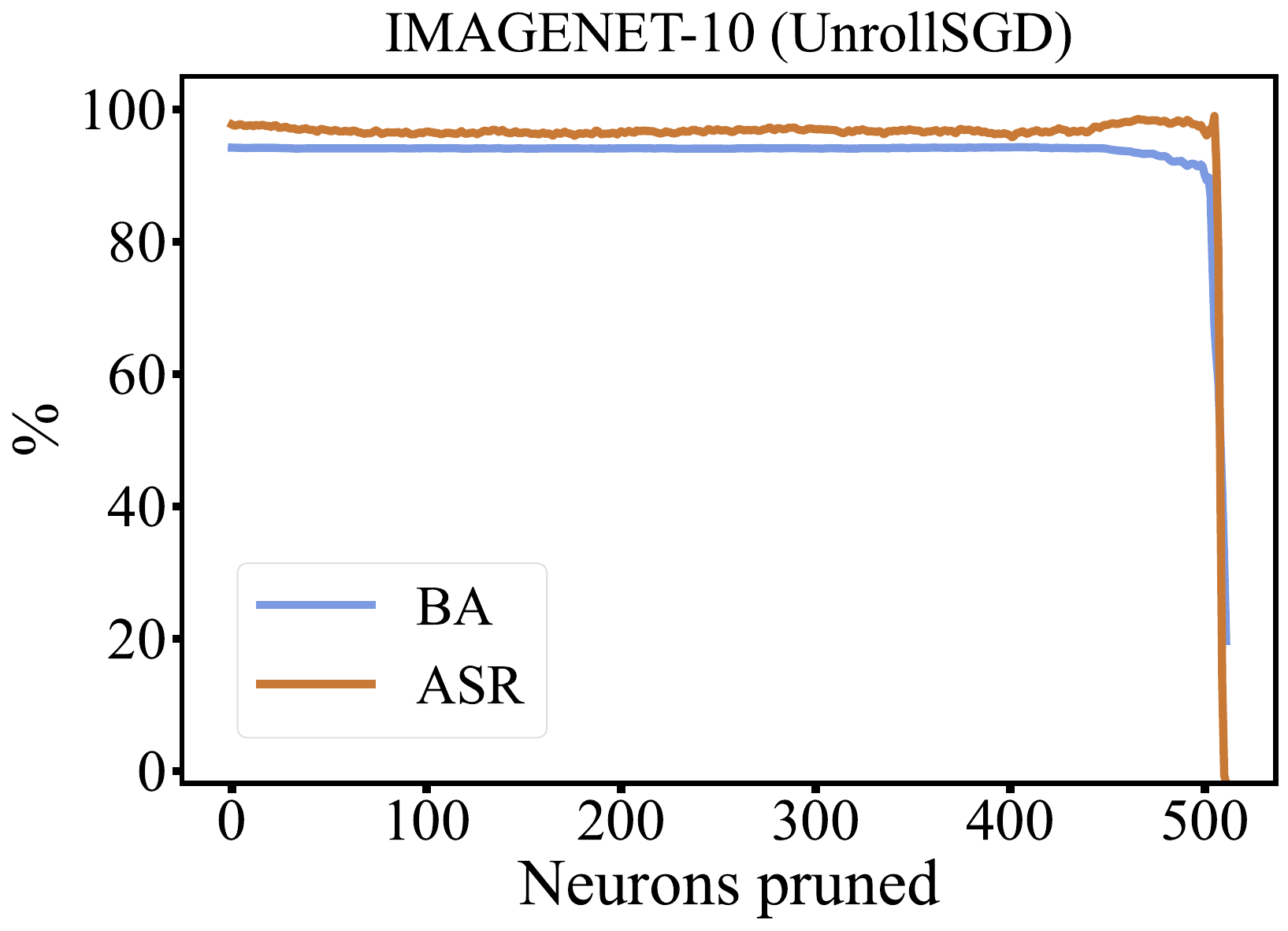}
  }
  \hspace{0.02\linewidth}
  \subfigure{
    \includegraphics[width=0.32\linewidth]{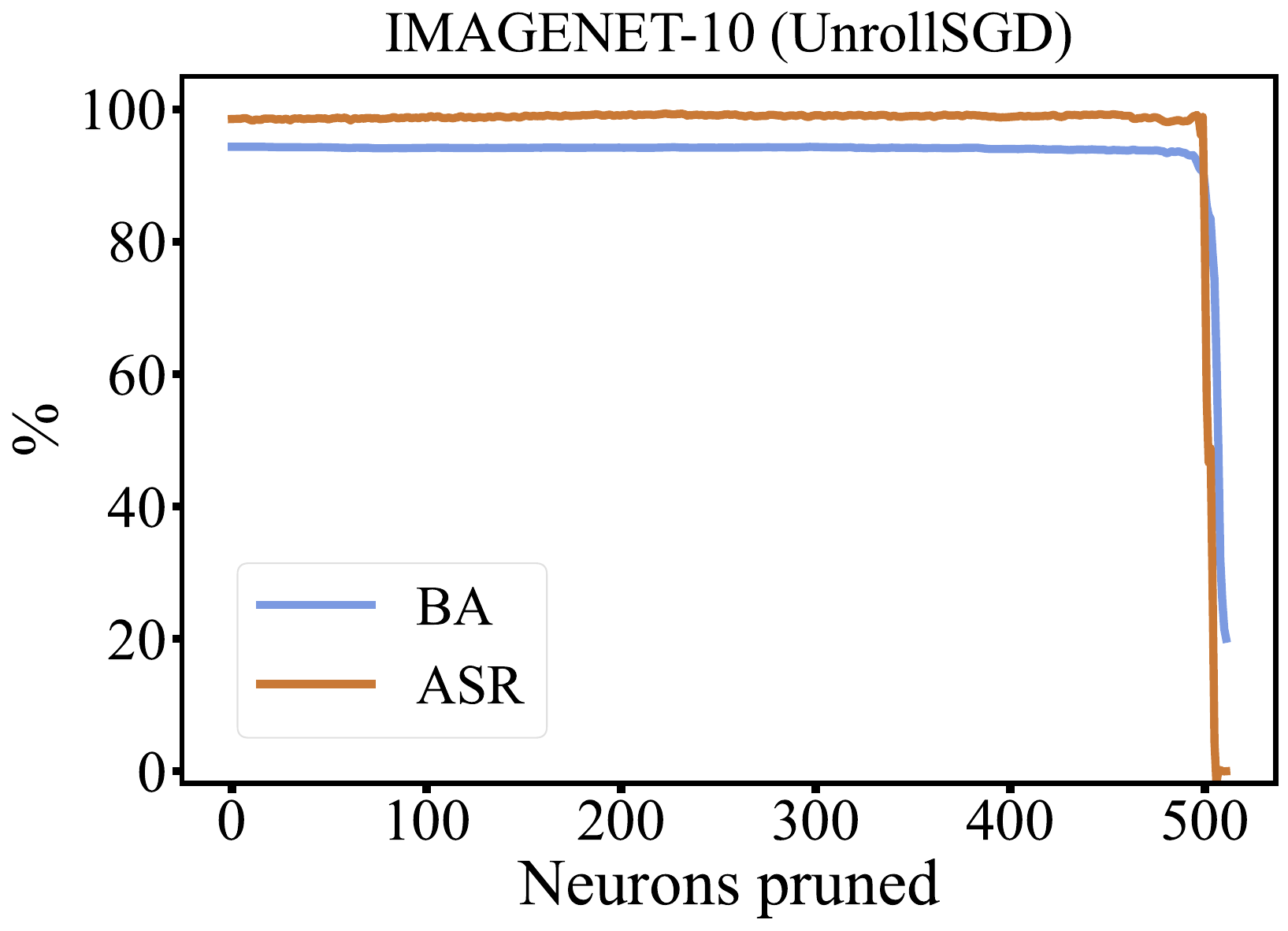}
  }\\[1.5ex]
  \subfigure{
    \includegraphics[width=0.32\linewidth]{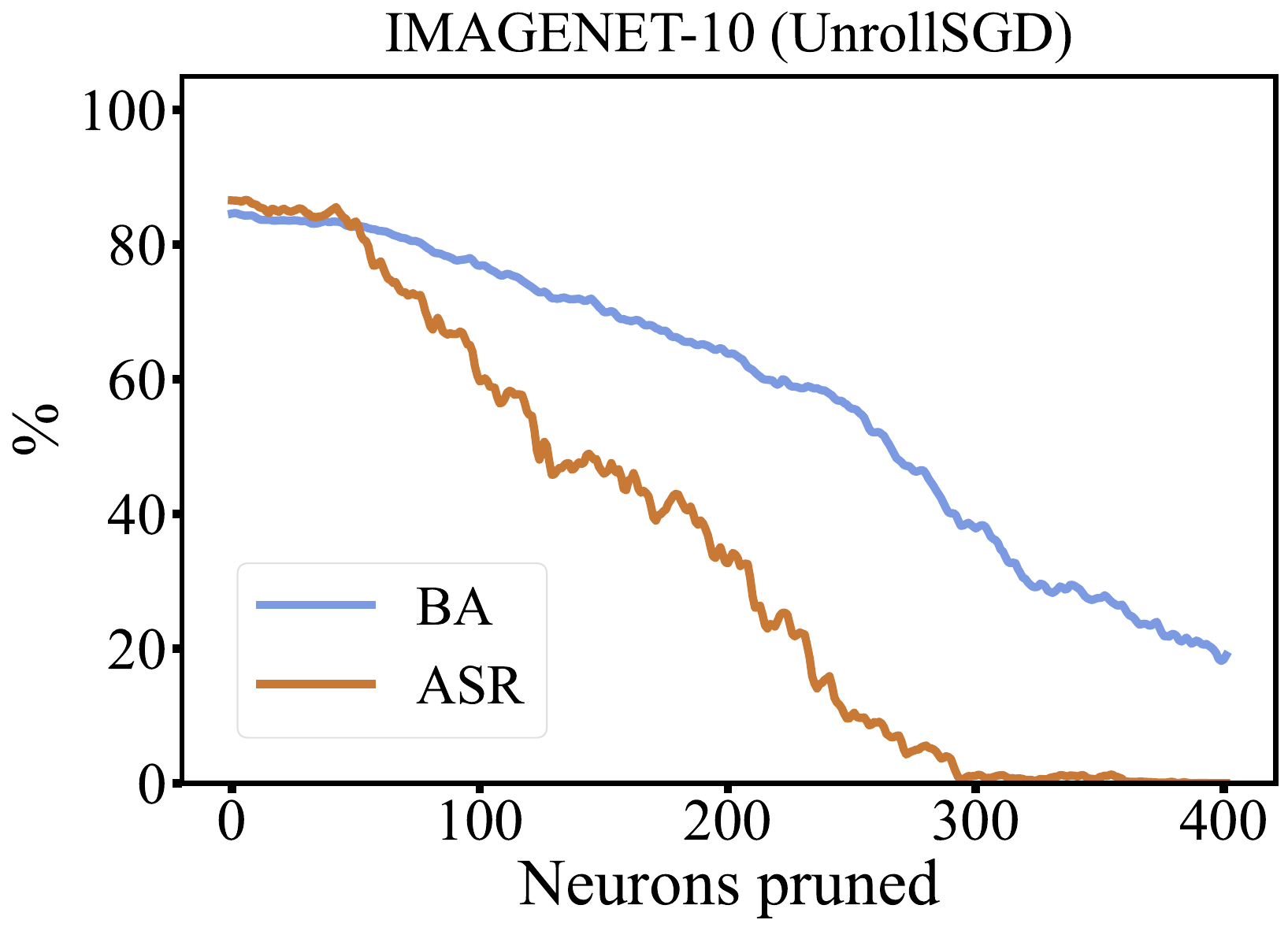}
  }
  \hspace{0.02\linewidth}
  \subfigure{
    \includegraphics[width=0.32\linewidth]{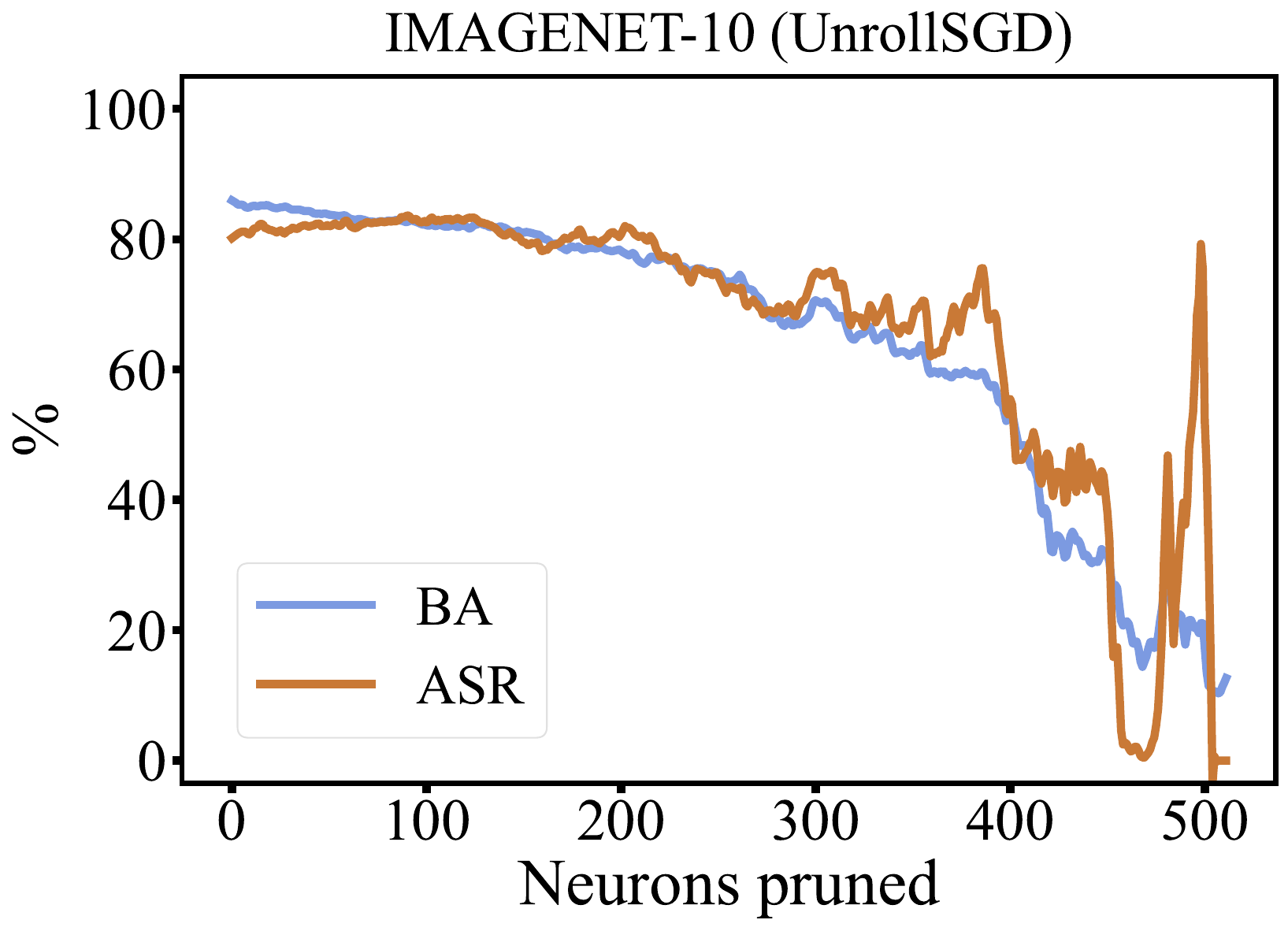}
  }
  \caption{Fine-pruning}
  \label{fig:fig1}
\end{figure*}

Fine-Pruning prunes neurons that are inactive on clean images, aiming to eliminate those responsible for backdoor behavior. We find that this defense cannot simultaneously maintain high clean accuracy and reduce attack success, indicating the resilience of our attack, as shown in Fig.~\ref{fig:fig1}.

\subsubsection{STRIP}
\begin{figure*}[t]
  \centering
  \subfigure{
    \includegraphics[width=0.32\linewidth]{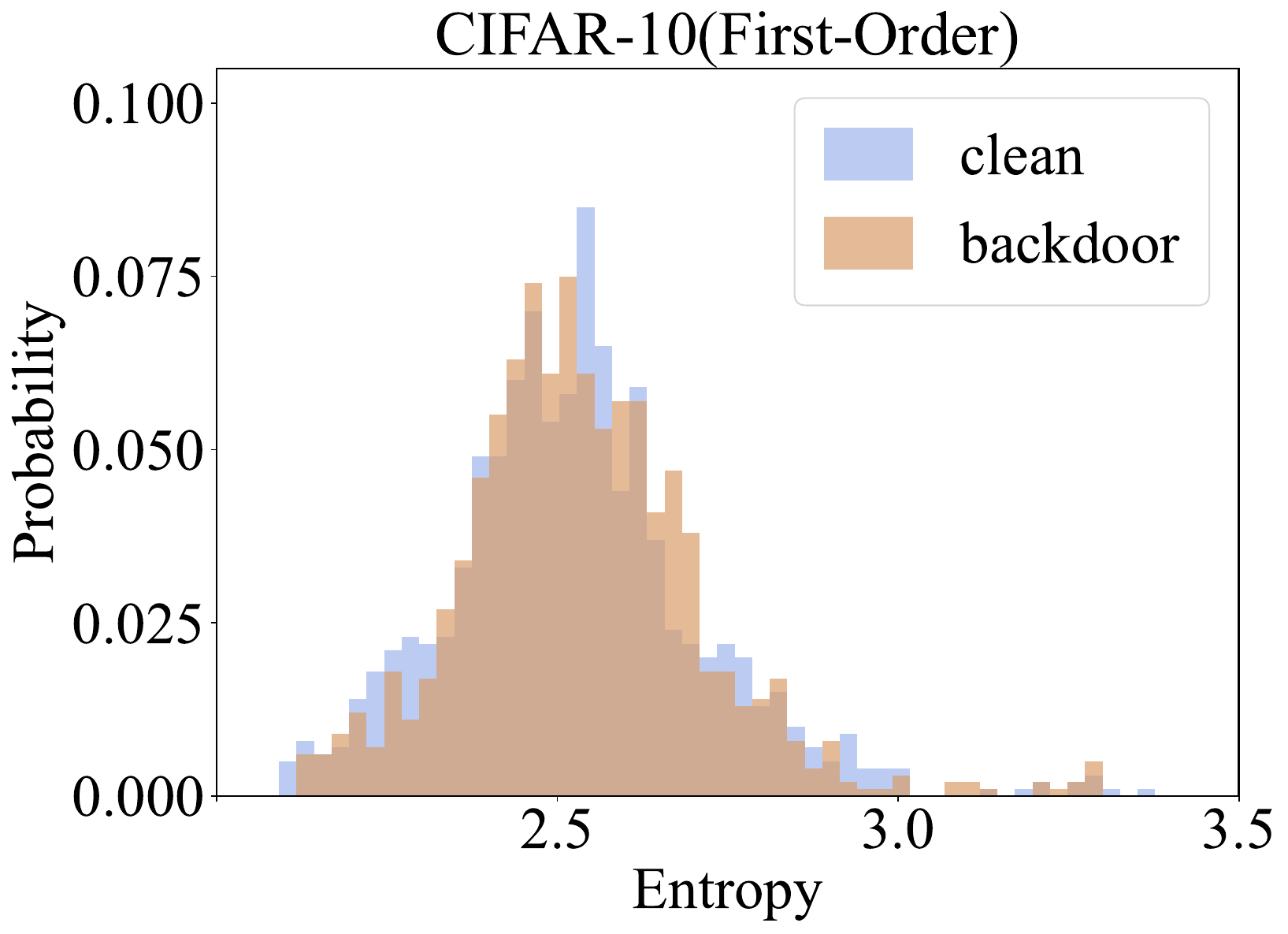}
  }
  \hspace{0.04\linewidth}  
  \subfigure{
    \includegraphics[width=0.32\linewidth]{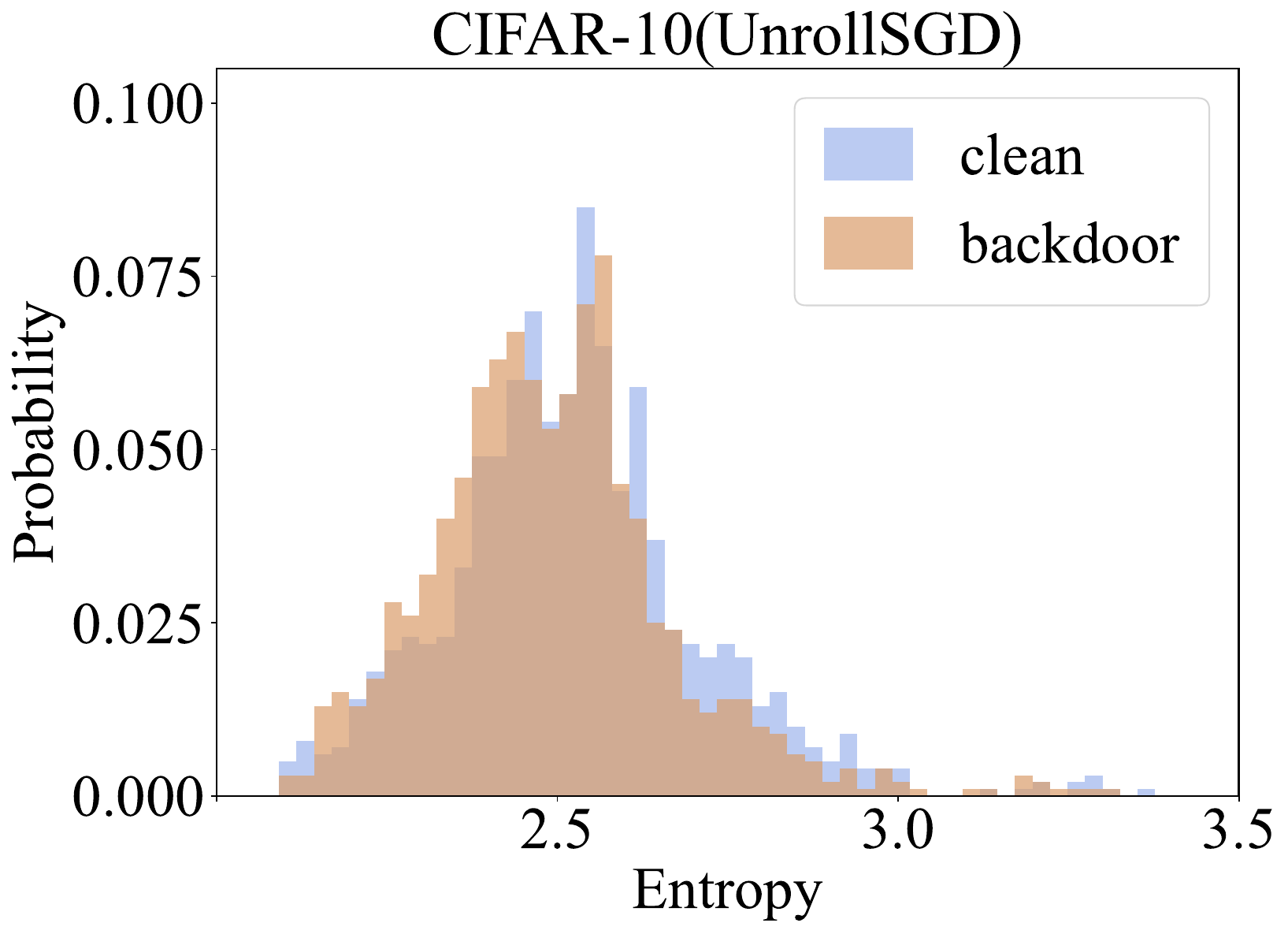}
  }\\[1.5ex]
  \subfigure{
    \includegraphics[width=0.32\linewidth]{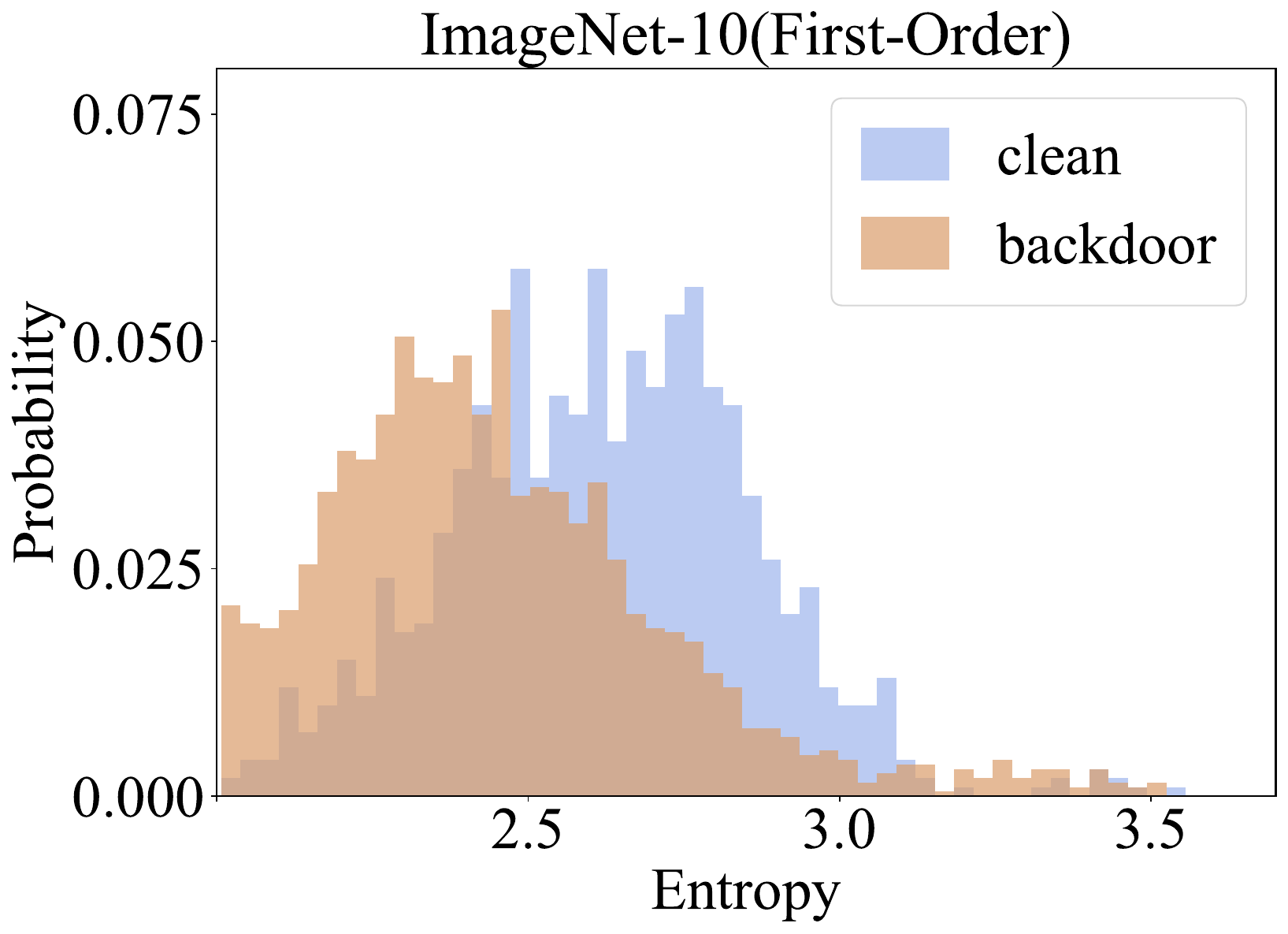}
  }
  \hspace{0.04\linewidth}
  \subfigure{
    \includegraphics[width=0.32\linewidth]{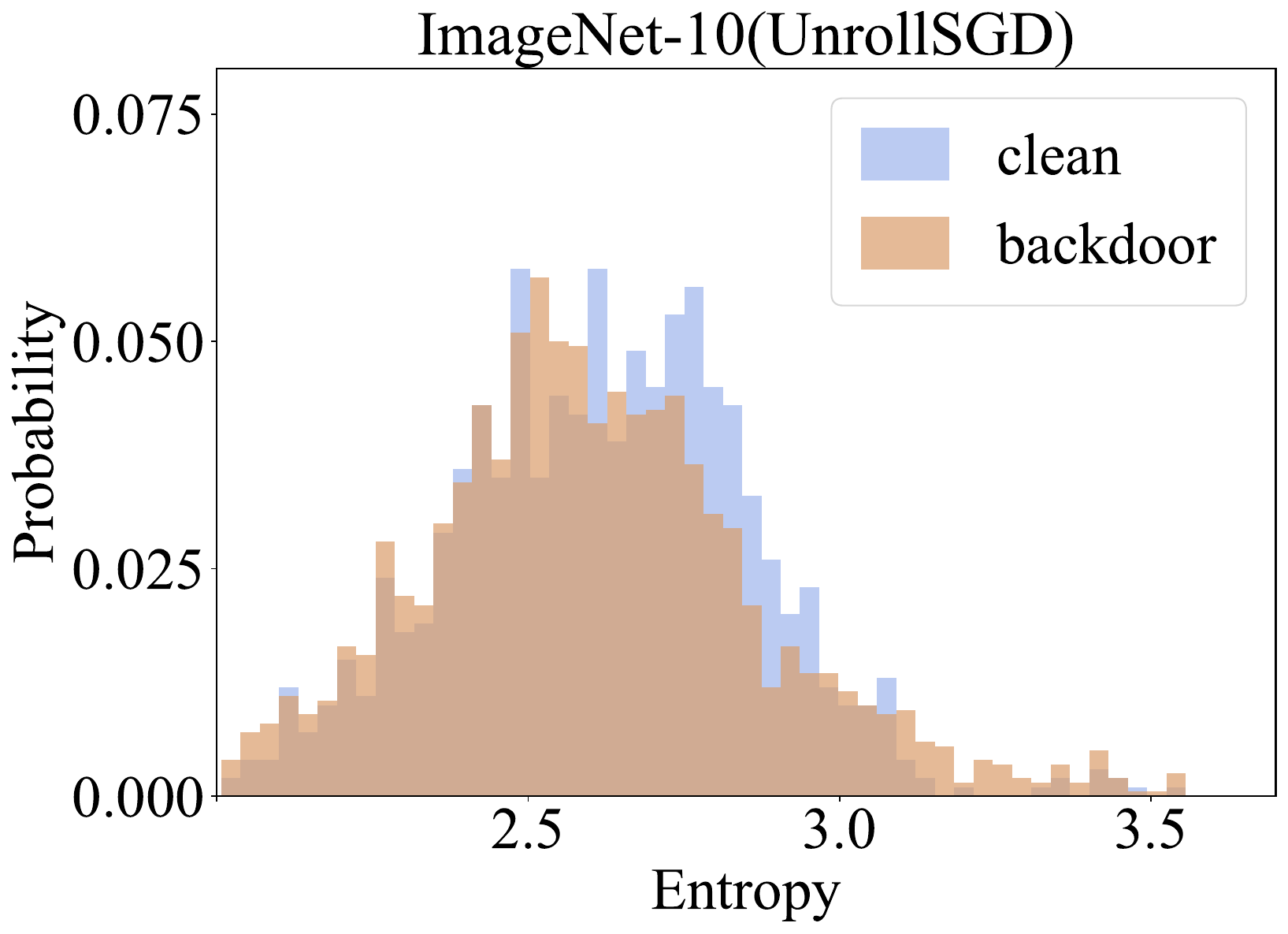}
  }
  \caption{STRIP}
  \label{fig:fig2}
\end{figure*}

STRIP detects potential backdoors by measuring the prediction consistency of perturbed inputs. Our attack produces entropy distributions similar to those of clean models, effectively bypassing this defense, as shown in Fig.~\ref{fig:fig2}.

\subsubsection{GradCAM-based Inspection}
\begin{figure*}[t]
    \centering
    \begin{minipage}[b]{0.32\linewidth}
        \centering
        \includegraphics[width=\linewidth]{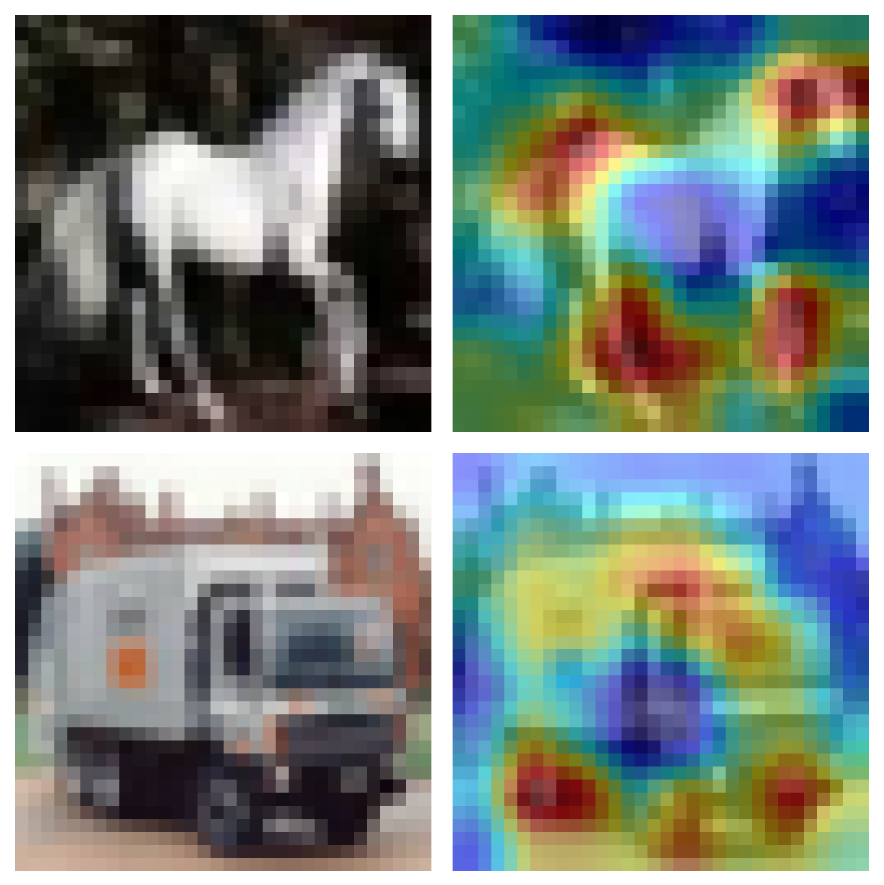}
        \caption*{(a) CIFAR-10 (Clean)}
    \end{minipage}
    \begin{minipage}[b]{0.32\linewidth}
        \centering
        \includegraphics[width=\linewidth]{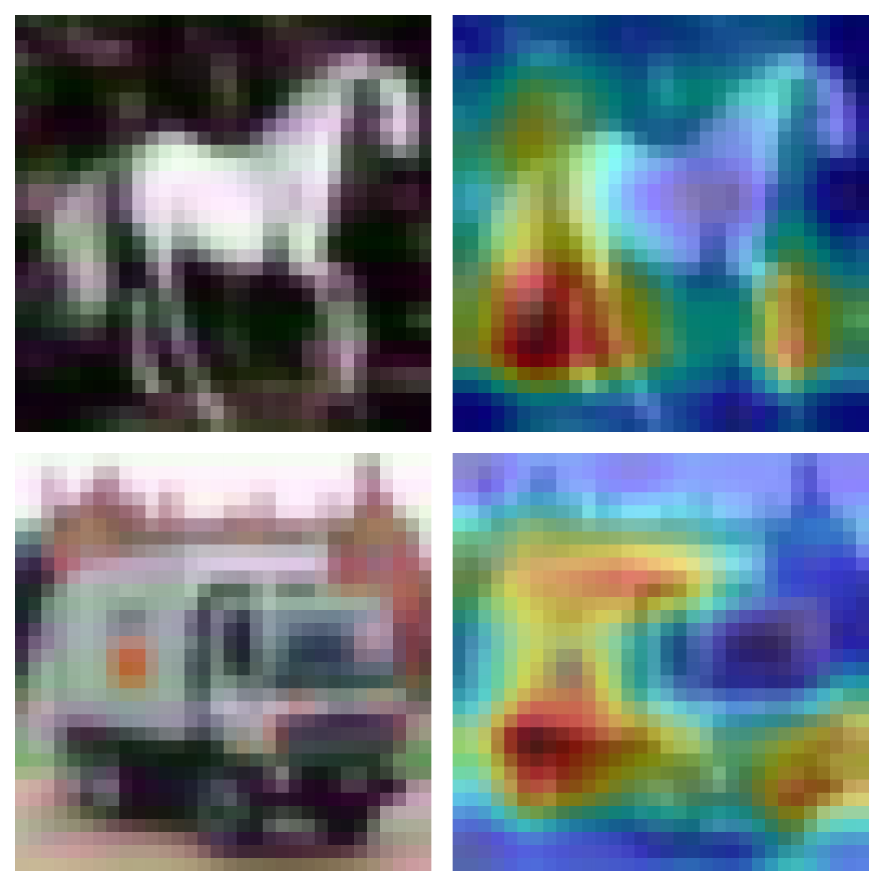}
        \caption*{(b) CIFAR-10 (First-Order)}
    \end{minipage}
    \begin{minipage}[b]{0.32\linewidth}
        \centering
        \includegraphics[width=\linewidth]{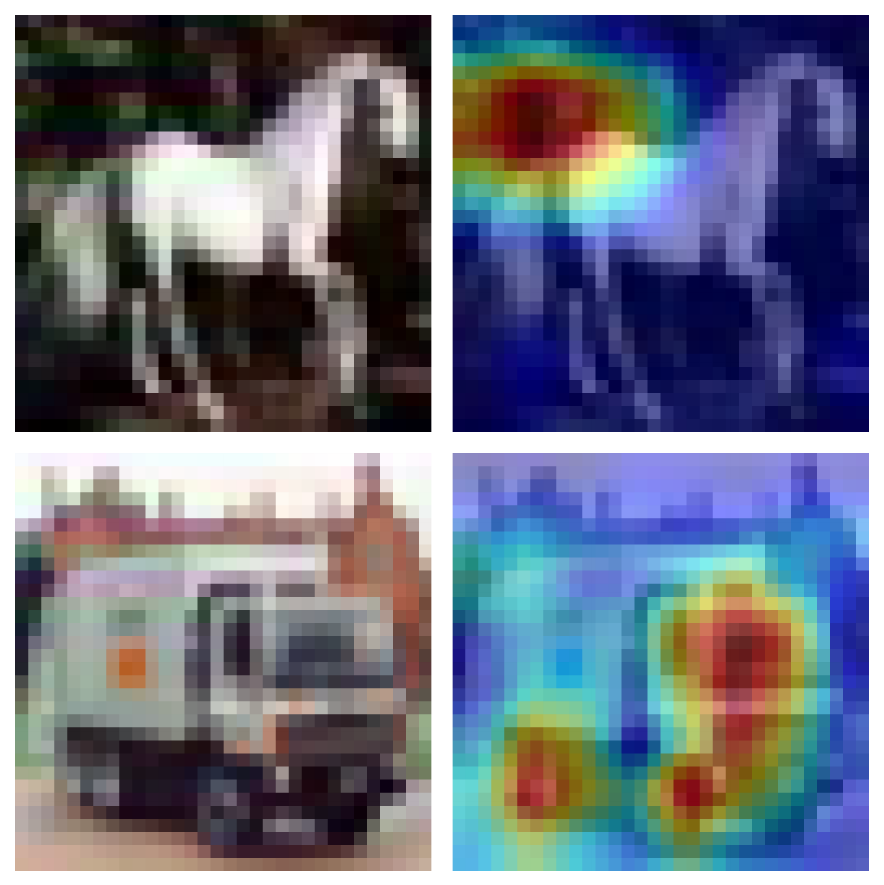}
        \caption*{(c) CIFAR-10 (UnrollSGD)}
    \end{minipage}
    \\[1.5ex]
    \begin{minipage}[b]{0.32\linewidth}
        \centering
        \includegraphics[width=\linewidth]{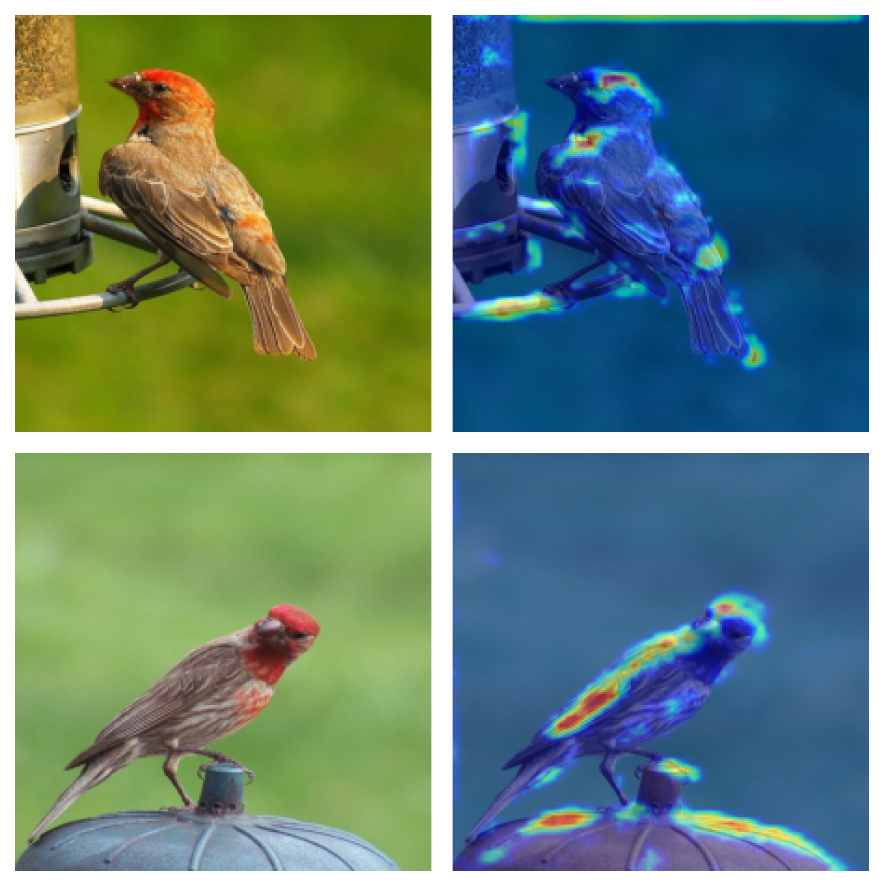}
        \caption*{(d) IMAGENET-10 (Clean)}
    \end{minipage}
    \begin{minipage}[b]{0.32\linewidth}
        \centering
        \includegraphics[width=\linewidth]{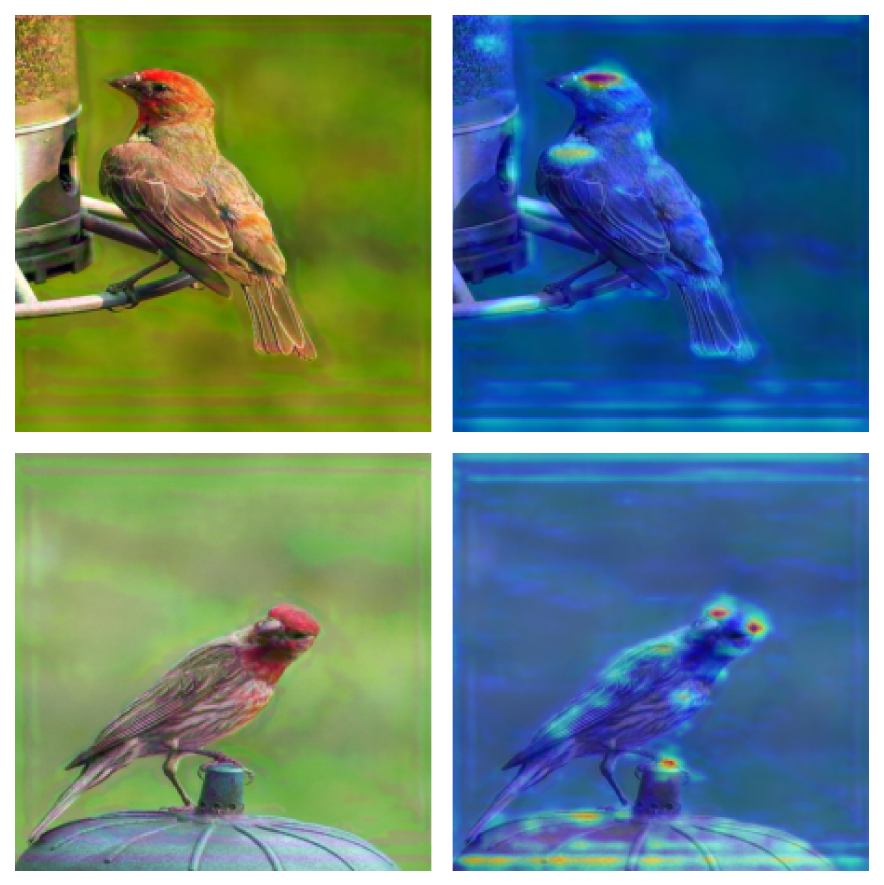}
        \caption*{(e) IMAGENET-10 (First-Order)}
    \end{minipage}
    \begin{minipage}[b]{0.32\linewidth}
        \centering
        \includegraphics[width=\linewidth]{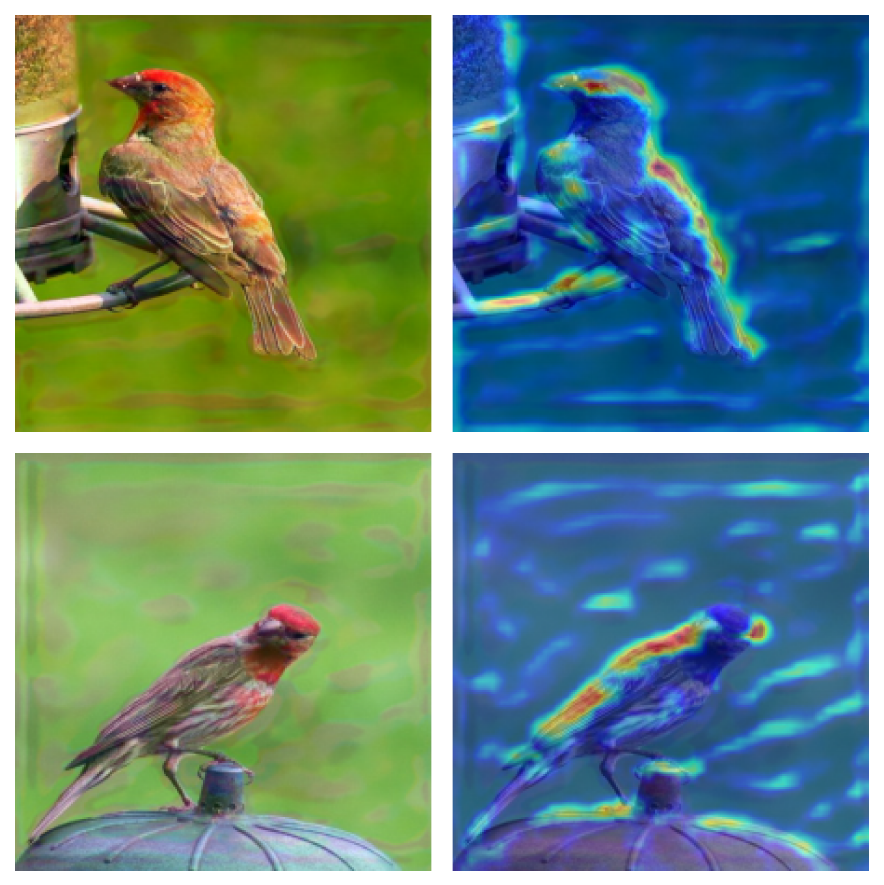}
        \caption*{(f) IMAGENET-10 (UnrollSGD)}
    \end{minipage}
    \caption{GradCAM visualization}
    \label{fig:fig3}
\end{figure*}

GradCAM highlights regions influencing model predictions. In our experiments, the heatmaps generated for poisoned models do not consistently reveal the trigger location, making our backdoor difficult to detect via visual inspection, as shown in Fig.~\ref{fig:fig3}.


\vfill

\end{document}